\documentclass[aps,pra,eqsecnum,twocolumn]{revtex4}
\usepackage{graphicx}
\usepackage{bm}
\usepackage{amsmath,amssymb,amsthm,dsfont,bm}
\usepackage{color}
\usepackage{wasysym}
\usepackage{ulem}
\usepackage{appendix}
\usepackage[applemac]{inputenc}
\usepackage[dvipsnames]{xcolor}

\begin{document}
\preprint{}
\title{Quantum metrology and coherence}
\author{Laura Ares}
\email{laurares@ucm.es}
\author{Alfredo Luis}
\email{alluis@fis.ucm.es}
\homepage{http://www.ucm.es/info/gioq}
\affiliation{Departamento de \'{O}ptica, Facultad de Ciencias
F\'{\i}sicas, Universidad Complutense, 28040 Madrid, Spain}
\date{\today}

\begin{abstract}
We address the relation between quantum metrological resolution and coherence. We examine this dependence in two manners: we develop a quantum Wiener-Kintchine theorem for a suitable model of quantum ruler, and we compute the Fisher information. The two methods have the virtue of including both the contributions of probe and measurement on an equal footing. We illustrate this approach with several examples of linear and nonlinear metrology. Finally, we optimize resolution regarding coherence as a finite resource. 
\end{abstract}
\maketitle

\section{Introduction}

Coherence is a basic and elusive concept in classical optics, as well as in modern quantum mechanics and quantum optics. Coherence is not a phenomenon by itself but it underlies the most relevant phenomena of optics and quantum mechanics, including current quantum technologies for which decoherence is the most fearful enemy \cite{decoh}. In recent times it has been recognized the fundamental role of coherence as a quantum resource for the most promising applications of the quantum theory \cite{rtc}. 

\bigskip                            

There have been many different theories formalising the concept of coherence and its practical consequences in quantum physics, from the earliest theory of Glauber, to the most recent in the form of quantum resource theories \cite{rtc,Glauber}. 

\bigskip

Among the areas of application of coherence we can include quantum metrology, in which the effort to perform the most careful observation of physical phenomena meets the limits imposed by the most bizarre consequences of quantum physics, involving basic issues such as state reduction and uncertainty relations \cite{qm}.

\bigskip

In this work we elaborate a fresh relation between quantum resolution and coherence to be derived from first principles. We propose a model of detection in which the measurement takes the form of a quantum ruler constructed by the same group of transformations that encodes the signal to be detected on the probe state. This leads to a resolution-coherence relation that fully mimics the classic Wiener-Kintchine theorem allowing a complete parallelism between quantum and classical coherence theories \cite{MW}. This approach is completed with the more usual formalism based on the quantum Cram\'er--Rao bound in terms of Fisher information.  We illustrate this approach by applying it to several relevant examples of linear and nonlinear metrology, looking as far as possible for a physical understanding of the results obtained \cite{nlm}.    

\bigskip

A distinctive feature of the coherence--resolution relation developed in this work is that it includes the quantum coherence properties of the detection process on an equal footing with the coherence of the probe state. We think this is relevant, specially regarding the optimization of resolution given the always limited finite amount of resources, where in the most common approaches only consider the resources consumed by the probe.

\bigskip

Our analysis confirms compelling evidence showing that the true resource for precise detection is coherence. In this spirit we look for the optimum resolution given the finite coherence resources conveyed by probe and measurement.  

\section{Quantum ruler}

We base most of our analysis on a rather natural and paradigmatic model of measurement, where the signal to detected shifts the pointer against a ticked ruler in which the ticks are actually repeated shifts of a basic mark. 

\bigskip

The pointer is a quantum state with density matrix $\rho_0$, which is transformed by a signal-dependent unitary transformation $D(\lambda)=\exp (-i \lambda G )$ to $\rho (\lambda) = D(\lambda) \rho_0 D^\dagger (\lambda)$, where $\lambda$ represents the signal to be detected, $G$ is the infinitesimal generator of the transformation, and we have $D^\dagger (\lambda) = D(- \lambda)$.

\bigskip

The ticked ruler is represented by the measurement performed, to be called $M$, to be described by a positive operator-valued measure $\Delta (m)$, where the outcomes $m$ represent the ticks of the ruler. We consider that the ticks are spaced by effect of the same transformation that shifts the probe, this is 
\begin{equation}
\label{si}
    \Delta (m)= D(m) \Delta_0 D^\dagger (m), \quad \Delta_0 = \Delta (0) ,
\end{equation}
for the same unitary transformation $D(m)$ above, and $\Delta_0$ is the origin for the ticks. For definiteness and without loss of generality we will assume that $m$ is a continuous variable extending from $-\infty$ to $\infty$.

\bigskip

Among other interesting properties, this structure guarantees that we can derive general conclusions about the performance of the detection process independent both the of the signal value $\lambda$ and of the outcome $m$ obtained. This is clearly manifest by computing the statistics of the measurement $p(m|\lambda)$ conditioned to the signal value $\lambda$, this is 
\begin{equation}
    p(m|\lambda )= \mathrm{tr} \left [ \rho (\lambda)  \Delta (m) \right ]  ,
\end{equation}
this is 
\begin{equation}
    p(m|\lambda ) =  \mathrm{tr} \left [ D(\lambda) \rho_0 D^\dagger (\lambda)  D(m) \Delta_0 D^\dagger (m)\right ] ,
\end{equation}
leading to 
\begin{equation}
\label{ff}
    p(m|\lambda ) = \mathrm{tr} \left [ \rho_0 D (m-\lambda)  \Delta_0 D^\dagger (m-\lambda)\right ] = p(m- \lambda ) .
\end{equation}
This conditional probability, along with the prior information we might have about $\lambda$, in the case there is any, is the building block of a complete signal-estimation process, for example within a Bayesian formulation \cite{JR19,DJK18}. 

A key result is that the conditional probability depends on $\lambda$ and $m$ just through the combination $m-\lambda$, this is $ p(m|\lambda ) = p(\mu )$, where $\mu = m- \lambda$. We can refer to this property as shift invariance \cite{si}. The performance of the whole detection process is given only by the form of $p(\mu)$ that depends just on  $\rho_0$, $\Delta_0$, and $G$. More specifically $p(\mu)$ is fixed by the coherence properties of  $\rho_0$ and $\Delta_0$ with respect to $G$, as we shall see clearly below. 

\bigskip

Before proceeding let us present a necessary condition that $\Delta_0$ must satisfy so that the shift-invariant ruler $\Delta (m)= D(m) \Delta_0 D^\dagger (m)$ provides a {\it bona fide} measurement. Given the fomr of $\Delta (m)$, the necessary and sufficient conditions for the reality, nonnegativity, and normalization of $p(\mu)$ are just nonnegativity of $\Delta_0$  and resolution of identity, this is 
\begin{equation}
\label{ir}
    \int dm \Delta (m)= I ,
\end{equation}
where $I$ is the identity. To show the consequences of condition (\ref{ir}) let us use the basis of eigenvectors of $G$, this is $G|g\rangle = g | g \rangle$, where $g$ will be assumed to be continuous and nondegenerate without loss of generality, with
\begin{equation}
    \int dg | g\rangle \langle g | = I.
\end{equation}
Then
\begin{equation}
    \int dm \Delta (m)= \int dm \int dg \int dg^\prime  | g\rangle \langle g | \Delta (m) | g^\prime \rangle \langle g^\prime | .
\end{equation}
Using Eq. (\ref{si}) and $D(m) | g \rangle = \exp(-i g m )| g \rangle$,
\begin{equation}
    \int dm \Delta (m)= \int dm \int dg \int dg^\prime e^{im(g-g^\prime )}  | g\rangle \langle g | \Delta_0 | g^\prime \rangle \langle g^\prime | ,
\end{equation}
so that performing the $m$ integration
\begin{equation}
    \int dm \Delta (m)= 2 \pi \int dg | g\rangle  \langle g | \Delta_0 | g \rangle \langle g | .
\end{equation}
The necessary and sufficient conditions that $\Delta_0$ must satisfy to provide a legitimate quantum ruler $\Delta (m)$  are 
\begin{equation}
\label{diag}
   \Delta_0 >0, \quad \langle g | \Delta_0 | g \rangle = \frac{1}{2\pi} , 
\end{equation}
which implies 
\begin{equation}
\langle g | \Delta (m) | g \rangle = \frac{1}{2\pi} .
\end{equation}
This condition is quite interesting since it means that $\Delta (m)$ are nontrivial just because of their nondiagonal terms in the generator basis. These are their coherence terms with respect to $G$, in the way coherence is understood in modern quantum mechanics. And to say coherence is to say nonclassicality. As a further consequence, condition (\ref{diag}) implies that the variance of $G$ in $\Delta (m)$ diverges.

\section{Quantum Wiener-Kintchine theorem}

Let us now address the form of $p(\mu)$ given its relation to $\rho_0$, $\Delta_0$, and $G$ in Eq. (\ref{ff}). We can proceed from Eq. (\ref{ff}) making use of the $G$ basis to get  
\begin{equation}
   p(\mu ) = \int dg \int d g^\prime
   \langle g |\rho_0 | g^\prime \rangle \langle g^\prime |\Delta_0 | g \rangle e^{i(g-g^\prime)\mu} .
\end{equation}
In order to highlight the relation between $p(\mu)$ and coherence, let us formally perform the change of variables $g^\prime = g + \tau$, to get
\begin{equation}
\label{WKt}
   p(\mu ) = \int d\tau \Gamma (\tau ) e^{-i\tau \mu} ,
\end{equation}
where 
\begin{equation}
\label{Gtau}
   \Gamma (\tau ) = \int^\prime d g 
   \langle g |\rho_0 | g + \tau \rangle \langle g + \tau |\Delta_0 | g \rangle  ,
\end{equation}
and by the prime we mean that the range of integration on $g$ may depend on $\tau$. Note that this relation naturally respects the conditions that $p(\mu)$ must satisfy as a probability density, this is to be real, positive, and normalized, so we have $\Gamma (0) = 1/(2 \pi)$ as well as $\Gamma^\ast (\tau) = \Gamma (- \tau)$.

\bigskip

We see that Eqs. (\ref{WKt}) and (\ref{Gtau}) provide a quantum exact replica {\it mutatis mutandis} of a celebrated theorem of classical coherence optics, the Wiener-Kintchine theorem \cite{MW,BW}, that establishes that the spectral density and the coherence function are a Fourier transform pair. In our case the statistics $p(\mu )$ plays the role of the spectral density while $\Gamma (\tau)$ plays the role of coherence function. 

We might better say that $\Gamma (\tau)$ is a detection-process coherence function made by the product of the individual probe and ruler coherence functions 
\begin{equation}
 \langle g |\rho_0 | g + \tau \rangle, \quad  \langle g |\Delta_0 | g + \tau \rangle ,
\end{equation}
respecting the fruitful equivalence between the matrix elements of the density matrix and a classical-optics coherence function. To ensure this coherence link let us note that the full dependence of $\Gamma (\tau)$ on $\tau$ relies just on the coherence terms, this is the nondiagonal terms of probe and ruler on the basis $G$, which is precisely the coherence basis that matters regarding metrology. 

\bigskip

As a rather relevant result of this coherence analysis, we highlight that the detection-process coherence function is fully symmetrical on the probe and the ruler, so the probe and ruler coherences contribute equally to the resolution. 

\bigskip

A final and interesting remark of this approach is that the coherence with respect to the observable $G$ is not the same that coherence with respect to the observable $G^2$, even though both $G$ and $G^2$ can be diagonal in the same basis, so that  the traditional resource theories will predict the same coherence. This is further illustrated in Appendix A.

\subsection{Resolution and coherence time}

The above theorem leads us to a natural relation between signal-detection resolution $\Delta \lambda$, understood as the width of the spectral-like function $p(\mu)$, and coherence time $\tau_c$,  understood as the width of the coherence-like function $\Gamma (\tau)$. 

\bigskip

To suitably establish their relation several approaches might be followed. Here we will follow the one presented in Ref. \cite{Perina} as particularly simple and insightful, which can be readily derived from the Parceval's relation
\begin{equation}
    2 \pi \int d \tau |\Gamma (\tau) |^2 = \int d \mu p^2 (\mu) .
\end{equation}
 Based on this we define the following measures of coherence and signal uncertainty, both relying in some version of Renyi entropies \cite{Renyi}. Regarding coherence let us consider  
 \begin{equation}
 \label{tc}
    \tau_c = \int d \tau |\gamma (\tau) |^2 ,
\end{equation}
where $\gamma (\tau)$ is the corresponding degree of coherence as the properly normalized coherence function
\begin{equation}
    \gamma (\tau) = \frac{\Gamma(\tau)}{\Gamma(0)}= 2 \pi \Gamma (\tau).
\end{equation}
Concerning resolution, we define the signal uncertainty $\Delta \lambda$ as
\begin{equation}
\label{WKR}
    \Delta \lambda = \frac{1}{2 \sqrt{\pi}\int d\mu p^2 (\mu)} .
\end{equation}
Combining both we get the resolution-coherence relation
\begin{equation}
    \label{rcrelation}
    \tau_c \Delta \lambda =  \sqrt{\pi} .
\end{equation}
So the signal uncertainty is inversely proportional to the coherence time. Note that the $\mu$ integration of $p(\mu)$ is normalized to one while the $\tau$ integration of $\Gamma (\tau)$ is not, so the two width measures look so different although they address the same idea.

\subsection{Gaussian model}

Let us elaborate further this relation with a particular Gaussian form for probe and ruler, say
\begin{equation}
\psi_0 (g) = \frac{1}{\sqrt{\Delta G \sqrt{2 \pi}}}  e^{-(g-g_0)^2/(4 \Delta^2 G)} e^{i k_0 g}  .
\end{equation}
where we assume the probe to be in a pure state $\rho_0 = |\psi_0 \rangle \langle \psi_0  |$, being $\psi_0 (g) = \langle g | \psi_0 \rangle$, and
\begin{equation}
\label{DGG}
    \langle g |\Delta_0 |g^\prime \rangle = \frac{1}{2\pi}  e^{-  \Delta^2 \Phi_M (g-g^\prime)^2/2} ,
\end{equation}
where  as before we refer  as $\Phi$ to the observable conjugate to $G$. Resorting once again to classical optics, quantum Gaussian states parallels Gaussian Schell-model sources \cite{MW,GS}. In these terms, the factor $\Delta \Phi_M$ actually expresses coherence in the degree of freedom expressed by $G$.

\bigskip

With all this we get that
\begin{equation}
\label{GG}
    \Gamma (\tau) = \frac{1}{2\pi}  e^{-(\Delta^2 \Phi_M+ \Delta^2 \Phi_S) \tau^2 /2} ,
\end{equation}
being 
\begin{equation}
    \Delta^2 \Phi_S  = \frac{1}{4\Delta^2 G} .
\end{equation}
This leads to a coherence time 
\begin{equation}
\label{Gt}
    \tau^2_c= \frac{\pi}{\Delta^2 \Phi_M+ \Delta^2 \Phi_S} ,
\end{equation}
and then, finally, the resolution becomes:
\begin{equation}
\label{Gl}
    \Delta^2 \lambda = \Delta^2 \Phi_M + \Delta^2 \Phi_S .
\end{equation}

\subsection{Fisher Information}
The results of this approach can be compared and complemented by the resolution lower limit given by the Cram\'er--Rao bound \cite{HC46,Fr99}
\begin{equation}
\label{CRB}
    \Delta^2 \lambda = \frac{1}{F},
\end{equation}
with Fisher information 
\begin{equation}
\label{Fi}
    F = \int dm \frac{1}{p(m|\lambda)} \left [ \frac{\partial p(m|\lambda)}{\partial \lambda} \right ]^2  = 
    \int d \mu \frac{1}{p(\mu)} \left [ \frac{\partial p(\mu)}{\partial \mu} \right ]^2    .
\end{equation}

\bigskip
Likewise, the Fisher information is bounded by above by the quantum Fisher information \cite{qFi}. For pure states, the quantum Fisher information becomes proportional to the variance on the probe state of the generator of the signal-dependent transformation,
\begin{equation}
    \label{QFI}
    F\leq F_Q=4 \Delta^2 G.
\end{equation}

\bigskip

As well as the previous one, this method also takes into account the resources conveyed by the detector. This enables us to distinguish between ideal, unrealistic measurements and more realistic measurements, where by ideal we mean that it does not contribute to the signal uncertainty, say $\Delta \Phi_M =0$. This may be crucial for further optimization of the resources involved.

\section{Linear metrology}
The above general analysis fits perfectly well to signal detection based on the Heisenberg-Weyl group of transformations. To this end let us consider the one-dimensional motion of a particle with position- and momentum-like operators satisfying the commutation relation $[X,P]=i$. This is equally valid for a one-mode field where $X$ and $P$ are the field quadratures. More specifically, let it be $G=P$ so we are detecting position shifts. 
 
 \bigskip
 
We will consider pure Gaussian states both for the probe and for the measurement. For the probe state $|\psi_0 \rangle$ we have in the position representation
\begin{equation}
\label{probe}
\psi_0 (x) = \frac{1}{\sqrt{\Delta X_S \sqrt{2 \pi}}} e^{-ip_0 x } e^{-(x-x_0)^2/(4 \Delta X_S^2)}  ,
\end{equation}
where as usual $\psi_0 (x) = \langle x |\psi_0 \rangle$ being $| x \rangle $  the eigenstates of position operator $X$, say $X| x \rangle = x| x \rangle $.

\bigskip

Concerning the measurement we consider statistics given by projection on  squeezed coherent states $|\varphi_{m,k} \rangle$ as being the displacement of a squeezed vacuum. In the position representation their wavefunction is 
\begin{equation}
\label{meas}
\varphi_{m,k} (x) = \frac{1}{\sqrt{\Delta X_M \sqrt{2 \pi}}}  e^{-ik x} e^{-(x-m)^2/(4 \Delta^2 X_M)}  ,
\end{equation}
being $\varphi_{m,k} (x) = \langle x |\varphi_{m,k} \rangle$. Note that we have the shift-invariant property granted in particular in the $m$ variables given that $|\varphi_{m,k} \rangle = D(m) |\varphi_{0,k} \rangle$. Since all occurs on the $x$-domain we can simplify matters if we get ride of the $k$ and $p_0$ dependences in the usual way of looking for the marginal for the $m$ outcome, so our POVM is 
\begin{equation}
\Delta (m) = \frac{1}{2\pi} \int dk |\varphi_{m,k} \rangle \langle \varphi_{m,k} | ,
\end{equation}
leading to a fully shift-invariant detector model $\Delta (m)= D(m) \Delta_0 D^\dagger (m)$ with 
\begin{equation}
\Delta_0 =  \frac{1}{2\pi} \int dk |\varphi_{0,k} \rangle \langle \varphi_{0,k} | ,
\end{equation}
leading to
\begin{equation}
    \Delta_0 = \frac{1}{\sqrt{2\pi} \Delta X_M} \int dx e^{-x^2/(2 \Delta^2 X_M)} | x \rangle \langle x |.
\end{equation}
In the $P$ representation we get  
\begin{equation}
\label{D0}
    \langle p |\Delta_0 |p^\prime \rangle = \frac{1}{2\pi}  e^{-\Delta^2 X_M (p-p^\prime)^2/2} ,
\end{equation}
in agreement conditions Eq. (\ref{diag}) and the Gaussian model (\ref{DGG}).

\bigskip

With all this we get as suitable particular examples of Eqs. (\ref{GG}), (\ref{Gt}), and (\ref{Gl})
\begin{equation}
    \Gamma (\tau) = \frac{1}{2\pi}  e^{-(\Delta^2 X_M+ \Delta^2 X_S ) \tau^2 /2} ,
\end{equation}
leading to a coherence time 
\begin{equation}
    \tau^2_c= \frac{\pi}{\Delta^2 X_M+ \Delta^2 X_S} ,
\end{equation}
and finally a signal uncertainty
\begin{equation}
\label{resP}
    \Delta^2 \lambda = \Delta^2 X_M + \Delta^2 X_S.
\end{equation}

\subsection{Cram\'er--Rao bound}
It is worth noting that $\Delta^2 \lambda$ coincides exactly with the Cram\'er--Rao bound 
\begin{equation}
    \Delta^2 \lambda = \frac{1}{F_P}= \frac{1}{ \frac{1}{\Delta^2 X_M + \Delta^2 X_S}}.
\end{equation}
whit $F_P$ being the Fisher information (\ref{Fi}).
This might be expected since  we can take the outcomes $m$ as a suitable estimator so that a Gaussian shift-invariant conditional distribution implies that the estimator is efficient \cite{DJK18}.

\bigskip

Summarizing, the larger the squeezing the lesser $\Delta X_{M,S}$, and thus the larger the coherence and the larger the resolution. This is a natural and intuitive relation, that nevertheless is not satisfied by other approaches to coherence where larger squeezing means lesser coherence as shown for example in Ref. \cite{AL10}.  Actually, Fisher information has already been proposed by itself as a suitable measure of coherence in Refs. \cite{AL12}.

\subsection{Optimization}

This resolution (\ref{resP}) can be further compared to the Cram\'er--Rao bound under an ideal $X$ measurement with projection-valued measure $\Delta (m) = |x=m\rangle \langle x=m| $, for the same Gaussian probe state (\ref{probe}). In such a case we obtain
\begin{equation}
\label{iP}
    \Delta^2 \lambda =\frac{1}{F_P}= \Delta^2 X_S, 
\end{equation}
with
\begin{equation}
    F_P = \frac{1}{\Delta^2 X_S} = 4 \Delta^2 P_S ,   
\end{equation}
where the last equality for $F_P$ shows that it is actually the quantum Fisher information (\ref{QFI}).

\bigskip

Comparing Eqs. (\ref{resP}) and (\ref{iP}) we can see the natural result that the ideal case is retrieved in the limit $\Delta X_M \rightarrow 0$. But it must be noted that this means an infinite amount of coherence or squeezing resources devoted to the measurement. This holds whatever the way resources are counted, this means either infinite coherence time, infinite nonclassicality, infinite squeezing, or even infinite energy. For example, in quantum optics the projection on the quadrature eigenstates $|x\rangle$ is carried out in an hodomdyne detection scheme only in the limit of a local oscillator much more intense that the system state being measured. This essentially means that in this  unrealistic limit all resources are devoted to the measurement, which is a fact normally no taken into account.

\bigskip

Therefore, it is worth examining the optimization in the case of finite resource including the resources employed in the measurement.

It is clear that the resource determining the resolution is the coherence expressed by the coherence times $\frac{1}{\Delta^2 X_M}$ and $\frac{1}{\Delta^2 X_S}$. So let us consider as resource some fixed amount of coherence split between probe and ruler so that the following quantity is held constant
\begin{equation}
\label{C}
  \frac{1}{\Delta^2 X_S}+\frac{1}{\Delta^2 X_M} = \mathrm{constant} = C.  
\end{equation}

It can be seen that the minimum of the signal uncertainty in Eq.(\ref{resP}) when $C$ is held constant holds when the split of the coherence is balanced between probe and ruler, this is $\frac{1}{\Delta^2 X_M} = \frac{1}{\Delta^2 X_S}$, giving a signal uncertainty just 2 times the ideal case in Eq. (\ref{iP}), this is 
\begin{equation}
    \Delta^{2} \lambda = 2 \Delta^{2}  X_S.
\end{equation}

\bigskip

We would like to point out that there is no relation whatsoever between resolution $\Delta \lambda$ and energy, since actually we have never specified the physical apparatus embodying the probe and ruler, i. e., this might be a free particle, or an harmonic oscillator, or anything else, so the Hamiltonian $H$ might be anything without altering a bit of the conclusions. This analysis shows that what really matters is the internal coherence structure of both probe and measurement. 

\section{Phase shifts}

Let us consider one of the most useful and studied generators. This is the free Hamiltonian of the harmonic oscillator, the number operator $N$
\begin{equation}
    G = N = \frac{1}{2} \left ( P^2 + X^2 \right ) ,
\end{equation}
which includes as the most relevant examples the free evolution of single mode fields. By the way, this is a balanced combination of the two basic nonlinear generators, we will come again to this later. In any case, the phase shift generated by $N$ is the fundamental basis of interference, which the most powerful detector as demonstrated in the detection of gravitational waves \cite{GW}.

\subsection{Quantum ruler}

There is a possibility to directly translate to this case the very general analysis made above in Sec. 4. To be  metrologically useful the probe states experiencing phase shifts must have a very large mean number of photons, which allows the useful approximation of the discrete spectrum of $N$ by a continuous one, extending its domain to the entire real axis as a good simplifying assumption. Similarly regarding the domain of variation for the phase. 

\bigskip

In this case, there is a readily physical picture of the observable $\Phi$ conjugate to $G$, This is the quantum-optical phase observable \cite{qop}. For a single-mode field $\Phi$ can be well described by the positive operator valued measure
\begin{equation}
    \label{POVMPH}
    \Delta (\phi ) =  |\phi \rangle \langle \phi |,
\end{equation}
where $ |\phi \rangle$ are the nonorthogonal, unnormalized Susskind-Glogower phase states
\begin{equation}
    |\phi \rangle = \frac{1}{\sqrt{2 \pi}} \sum_{n=0}^\infty e^{-i\phi n} | n \rangle,
\end{equation}
being $| n \rangle$ the photon-number states as the eigenstates of the number operator $N$, this is $N| n \rangle = n | n \rangle$. There is clearly shift invariance, since
\begin{equation}
    \Delta (\phi ) = D(\phi) \Delta_0  D^\dagger (\phi) ,
\end{equation}
with 
\begin{equation}
    D(\phi ) = e^{-i \phi N}, \quad \Delta_0 = \Delta (\phi = 0) .
\end{equation}
This represents the case of an unrealistic phase measurement. To deal with a more realistic one we just replace $\Delta_0 $ by
\begin{equation}
    \label{GMP}
    \Delta_0 = \frac{1}{\sqrt{2\pi} \Delta \Phi_M} \int d\phi e^{-\phi^2/(2 \Delta^2 \Phi_M)} | \phi \rangle \langle \phi | ,
\end{equation}
leading in the number representation to
\begin{equation}
    \langle n |\Delta_0 |n^\prime \rangle = \frac{1}{2\pi}  e^{-\Delta^2 \Phi_M (n-n^\prime)^2/2} ,
\end{equation}
and we have used the above-mentioned approximation for the variable $n$. Furthermore, we can consider that the probe may be described by a Gaussian in the number basis as
\begin{equation}
\label{GPP}
\psi_0 (n) \simeq \frac{1}{\sqrt{\Delta N_S \sqrt{2 \pi}}}  e^{-(n-\bar{n})^2/(4 \Delta N_S^2)}  .
\end{equation}

\bigskip

With all this we get that
\begin{equation}
    \Gamma (\tau) \simeq \frac{1}{2\pi}  e^{-(\Delta^2 \Phi_M+ \Delta^2 \Phi_S ) \tau^2 /2} ,
\end{equation}
where 
\begin{equation}
    \Delta^2 \Phi_S \simeq \frac{1}{4 \Delta^2 N_S}.
\end{equation}
This leads to a coherence time 
\begin{equation}
    \tau^2_c= \frac{\pi}{\Delta^2 \Phi_M+ \Delta^2 \Phi_S} ,
\end{equation}
and then finally
\begin{equation}
\label{NPh}
    \Delta^2 \lambda = \Delta^2 \Phi_M + \Delta^2 \Phi_S ,
\end{equation}
which clearly reproduces the structure of preceding results.

\subsection{Cram\'er--Rao bound }

For the same settings, this is, a Gaussian probe state (\ref{GPP}) and a  unrealistic, Gaussian, measurement (\ref{GMP}), the obtained uncertainty (\ref{NPh}) coincides exactly with the Cram\'er--Rao bound calculated by (\ref{CRB}) and (\ref{Fi}),
\begin{equation}
    \Delta^2 \lambda = \Delta^2 \Phi_M + \Delta^2 \Phi_S .
\end{equation}

In the case of unrealistic phase measurement, $\Delta(\phi)=|\phi\rangle\langle\phi|$, the only contribution that remains is the uncertainty of the probe state, $\Delta^2 \lambda=\Delta^2\Phi_S $. As in the linear case, this uncertainty coincides with the inverse of the quantum Fisher information.

\subsection{Coherent-squeezed scheme}

It can be interesting to analyze the performance of the scheme introduced in the preceding section in terms of the more accessible settings regarding the probe state and the linear ruler in Eqs. (\ref{probe}) and (\ref{meas}), respectively. 

\bigskip

\begin{widetext}
In this case, the Fisher information can be computed using the good transformation properties of Wigner function. At $\lambda =0$, which may properly account for the case of small enough signal values, it leads to
\begin{equation}
\label{FN}
F_N = \frac{\left ( \Delta^2 X_S - \Delta^2 P_S \right)^2}{\left ( \Delta^2 X_S+\Delta^2 X_M \right )\left (\Delta^2 P_S+\Delta^2 P_M \right )} +  \frac{x_0^2}{\Delta^2 P_S+\Delta^2 P_M} + \frac{p_0^2}{\Delta^2 X_S + \Delta^2 X_M} .
\end{equation}

\bigskip

The terms depending on the displacements $x_0$ and $p_0$ reproduce the structure that we found in the following section devoted to the nonlinear case $G=P^2$. 

\end{widetext}
\subsection{Resolution as phase uncertainty}
Let us show that the Fisher information (\ref{FN}) can be fully expressed in terms of phase uncertainty. We find this quit suggestive since phase fluctuations are a key tool to understand coherence in the classical domain.

\bigskip

To this end we can join $\Delta X_M$ to $\Delta X_S$ as being additional effective uncertainty caused by some kind of blurry origin as
\begin{equation}
    \Delta^2 \tilde{X}_S = \Delta^2 X_S + \Delta^2 X_M .
\end{equation}
So the combination $\Delta \tilde{P}_S / x_0$ is the phase uncertainty for a state centered at $x_0$ with a $\Delta \tilde{P}_S$ uncertainty along $Y$. Similarly for the term $\Delta \tilde{X}_S /p_0$. So the last two terms are determined for phase uncertainty related to the displacement term in Eq. (\ref{FN}).

\bigskip

On the other hand, the first term in Eq. (\ref{FN}) is independent of the displacements $x_0$ and $p_0$, so it can be suitably understood as phase uncertainty for an squeezed vacuum. We can address this quadrature-based phase uncertainty for the vacuum in terms of its Wigner function for the squeezed vacuum $W_S (x,p)$ where the coordinates $x,p$ refer to the blurry variables $\tilde{X}_S, \tilde{P}_S$ including the fluctuations added by the detection process. After changing to polar coordinates  $x=r \cos \phi$, $p =r \sin \phi$ and integrating on $r$ to get a phase distribution $W_S (\phi)$ 
\begin{equation}
   W_S (\phi)\propto \frac{1}{1+\left ( \frac{1}{\Delta^2 \tilde{X}_S} - \frac{1}{\Delta^2 
   \tilde{P}_S} \right ) \sin^2 \phi} .
\end{equation}
This phase distribution very much recalls the resolution of a Fabry-Perot interferometer \cite{BW} suggesting a phase uncertainty of the form 
\begin{equation}
    \Delta^2 \phi_0  \simeq  \frac{1}{\left |\frac{1}{\Delta^2 \tilde{X}_S} - \frac{1}{\Delta^2 \tilde{P}_S} \right |}= \frac{\Delta^2 \tilde{X}_S \Delta^2 \tilde{P}_S}{\left | \Delta^2 \tilde{X}_S -\Delta^2 \tilde{P}_S \right |}.
\end{equation}

\bigskip
So, {\it grosso modo} we have that the Fisher information for phase shifts is fully expressible in terms of phase uncertainties as
\begin{equation}
    F_N \propto \frac{1}{\Delta^2 \phi_0}+\frac{1}{\Delta^2 \phi_{x}}+\frac{1}{\Delta^2 \phi_{p}}.
\end{equation}

\subsection{Non-Gaussian scenario}
To conclude, we employ the phase shift detection to test our quantum Wiener-Kintchine theorem when applied to non-Gaussian states. To this end, we consider an ideal phase measurement (\ref{POVMPH}) over a normalizable version of the Susskind-Glogower phase states as the probe state,
\begin{equation}
  |\xi\rangle =\sqrt{1-|\xi|^2}\sum_{n=0}^\infty \xi^n |n\rangle,
\end{equation}
specially interesting when $|\xi| \rightarrow1$. 

The signal uncertainty obtained by computing (\ref{WKR}) is

\begin{equation}
   \Delta^2 \lambda =\pi \left ( \frac{1-|\xi|^2}{1+|\xi|^2} \right )^2.
\end{equation}

\bigskip

We compare this result with the arisen from calculating the Fisher information in (\ref{Fi}), which conduces to

\begin{equation}
    \Delta^2 \lambda =\frac{(1 - |\xi|^2)^2}{2 |\xi|^2}.
\end{equation}

It is worth noting the similarity between the findings of both methods, regarding that we are working with $ |\xi| \rightarrow1$. In this limit, the quantum Wiener--Kintchine theorem gives a resolution of $\pi(1 - |\xi|^2)^2 $ and the Cram\'er--Rao bound of $2(1 - |\xi|^2)^2 $.
\bigskip

This example shows the generality of the theorem developed in Sec. III beyond the more developed Gaussian model. 

\section{Nonlinear metrology}

By nonlinear metrology we refer to generators $G$ beyond the Heisenberg-Weyl group of transformations, for example let $G=P^2$. 

\bigskip
In this case, the rather accessible settings already used above for the linear metrology, with Gaussian forms for the probe and the measurement, do not satisfy the invariance condition in (\ref{diag}). The general result obtained in (\ref{rcrelation}), still valid, would need a different physical implementation which respects the invariance condition.
\subsection{Cram\'er--Rao bound}
 
Nevertheless, in order to extend our analysis of the relation between resolution and coherence to the nonlinear case, we make the ansatz supported by the preceding examples that the Fisher information provides a tool compatible with the approach presented above. Thus, we analyze the performance of nonlinear detection via the Fisher Information regarding the probe state and the linear ruler in the form specified in Eqs. (\ref{probe}) and (\ref{meas}), respectively. This is we consider probe and ruler being just plain coherent-squeezed states and we relate the resolution to the squeezing as done in the linear case . We must take into account that now the $k$ outcomes are not trivial and contribute to the statistics.

\bigskip

Then, we estimate the resolution via the Cram\'er--Rao bound 
\begin{equation}
    \Delta^2 \lambda = \frac{1}{F_{P^2}}.
\end{equation}
The Fisher information evaluated at $\lambda =0$, becomes
\begin{equation}
\label{FP2}
 F_{P^2}=\frac{\Delta^2 P_S}{\Delta^2 P_M}
F_P^2+4 p_0^2 F_P,
\end{equation}
where $\Delta P_{S,M}$ represent the uncertainties of the operator $P$ in the probe and measurement states  
\begin{equation}
\label{ur}
     \Delta P_{S,M} = \frac{1}{2 \Delta X_{S,M}} ,
\end{equation}
and $F_P$ is the Fisher information corresponding to the linear case, $G=P$,
\begin{equation}
\label{FL}
    F_P=\frac{1}{\Delta^2X_0+\Delta^2 X_M} .
\end{equation}
\bigskip

Let us compare again the obtained Fisher information (\ref{FP2}) with the Fisher information in the case of an  unrealistic measurement of the observable $X$ after the action of the nonlinear signal-induced transformation generated by $G=P^2$.
At $\lambda = 0$ we obtain
\begin{equation}
 F_{P^2}=16 p_0^2\Delta^2 P_S ,
\end{equation}
which coincides with the limit $\Delta^2 X_M\rightarrow 0$ in Eqs (\ref{FP2}), (\ref{ur}) and (\ref{FL}). 

\subsection{Linear and nonlinear terms}

There are two well differentiated contributions to the Fisher information $F_{P^2}$ in Eq. (\ref{FP2}). This splitting can be easily understood is we transfer the displacement $p_0$ from the probe state to the transformation, this is 
\begin{equation}
    P^2 \rightarrow \left (P+p_0 \right )^2 = P^2 + 2 p_0 P + p_0^2 ,
\end{equation}
where the transformation generated by the right-hand side observable is acting on a probe with vanishing mean momentum.
The constant term $p_0^2$ is trivial regarding transformations since it produces no effect. So we get the combination of a nonlinear transformation $P^2$ and a linear transformation $2 p_0 P$ acting on a probe with vanishing mean momentum. In the linear part  $2 p_0 P$ we can observe that the momentum displacement $p_0$ acts amplifying the signal value $\lambda$ by a factor $2p_0$. Since we already know well about the linear part let us develop further the first nonlinear term in Eq. (\ref{FP2}) considering $p_0=0$ 
\begin{equation}
 F_{P^2}=\frac{\Delta^2 P_S}{\Delta^2 P_M}
F_P^2 ,
\end{equation}
expressing it in terms of $X$ variances using $\Delta X \Delta P = 1/2$ to get 
\begin{equation}
\label{nlp}
 F_{P^2}=
 \frac{\Delta^2 X_M/\Delta^2 X_S}{\left (\Delta^2 X_M + \Delta^2 X_S \right )^2} ,
 \end{equation}
 and finally
 \begin{equation}
 \label{ft}
 F_{P^2}= 
 \frac{2 \Delta^2 X_M \Delta^2 X_S}{\left (  \Delta^2 X_M + \Delta^2 X_S \right )^2} QF_{P^2}
\end{equation}
where $QF_{P^2}$ is the quantum Fisher information of the probe  
\begin{equation}
    QF_{P^2}= 4 \Delta^2 G = \frac{1}{2 \left ( \Delta^2 X_S \right )^2} .
\end{equation}

\subsection{Optimization nonlinear term}

The first factor in Eq. (\ref{ft}) depends symmetrically on the probe and measurement as in the linear case. For a fixed joint amount of coherence (\ref{C}), the optimum value of this nonlinear term is obtained when 
\begin{equation}
\frac{1}{\Delta^2 X_S} = 3\frac{1}{\Delta^2 X_M}, 
\end{equation}
leading for this contribution to $F_{P^2} = \frac{3}{8}QF_{P^2}$.

\section{Conclusions}

We have developed a theory of quantum coherence suited to be applied to quantum metrology. This formulation shows via a simple model that coherence is the true resource behind resolution, as suitably expressed by a quantum version of the Wiener--Kintchine theorem. This is also confirmed by the more common Fisher information analysis. 

One of the virtues of this formalism is that it shows that both the apparatus and probe coherence properties contribute equally to the detection performance. This is a valuable result that acknowledges that system and detector are inextricably mingled to produce the observed statistics as required by basic quantum postulates such as the very Born's rule. 

\bigskip
\appendix
\section{Coherence relative to $G^2$ and $G$ are different}

As commented within the body of the manuscript, a really interesting property of the theory introduced in this work is that the coherence with respect to the observable $G$ is not the same that coherence with respect to the observable $G^2$, even though both $G$ and $G^2$ can be diagonal in the same basis, so that  the traditional resource theories will predict the same coherence. 

\bigskip

To show this let us consider $G_1=P$ and $G_2 = P^2 = G_1^2$ within a model of the most simple case of a pure Gaussian state for the probe
\begin{equation}
\psi_0 (p) = \frac{1}{\sqrt{\Delta P \sqrt{2 \pi}}} e^{-(p-p_0)^2/(4 \Delta^2 P)}  .
\end{equation}
Let us focus on a simplified version of the coherence function just missing the apparatus part for simplicity. For the case $G_1=P$ we get
\begin{equation}
\label{G1}
   \Gamma_1 (\tau ) = \int^\prime d p 
   \langle p |\rho_0 | p + \tau \rangle = e^{-\tau^2/(8\Delta^2 P )} .
\end{equation}
leading to a coherence time  $\tau_c \propto \Delta P$ independent of $p_0$. In the case  $G_2=P^2 = G_1^2$  we get, for $p_0=0$
\begin{equation}
\label{G2}
   \Gamma_2 (\tau ) = \int^\prime d p 
   \langle p |\rho_0 | \sqrt{p^2 + \tau} \rangle = e^{-|\tau| /(4\Delta^2 P )},
\end{equation}
leading to a coherence time  $\tau_c \propto \Delta^2 P$. Moreover it can be seen numerically that when $p_0 \neq 0$ the result clearly depends on $p_0$. 

\bigskip

It might be argued that the main difference between coherence function (\ref{G1}) and (\ref{G2}) is a matter of nomenclature and that the results are equivalent provided we replace $\tau$ by $\tau^2$ in Eq. (\ref{G2}). However, this is not quite so, since in such a case we should also replace $\tau$ by by $\tau^2$ in the Wiener-Kintchine theorem (\ref{WKt}), so that the final statistics is the same using either $\tau$ or $\tau^2$. In other words, the $\tau$ differences between Eqs. (\ref{G1}) and (\ref{G2}) are fully meaningful in the sense that they express the different signal resolutions provided by the linear and nonlinear schemes. 

\bigskip

\noindent{\bf Acknowledgments.- }
L. A. and A. L. acknowledge financial support from Spanish Ministerio de Econom\'ia y Competitividad Project No. FIS2016-75199-P.
L. A. acknowledges financial support from European Social Fund and the Spanish Ministerio de Ciencia Innovaci\'{o}n y Universidades, Contract Grant No. BES-2017-081942. We thank Dr. J. Rubio for continuous support and helpful comments.

\end{document}